\documentclass[12pt]{article}

\usepackage{graphicx,graphics}

\usepackage{fullpage}
\usepackage{amssymb}

\usepackage{amsmath}

\usepackage{color}
\usepackage{natbib}
\usepackage{algorithm}
\usepackage{algorithmic}

\usepackage{url}

\usepackage{color} 
\usepackage{listings}

\def\ga{\mbox{Ga}}
\def\iga{\mbox{IGa}}
\def\be{\mbox{Be}}
\def\ber{\mbox{Ber}}

\def\no{\mbox{N}}

\def\un{\mbox{Un}}

\def\dir{\mbox{Dir}}

\def\mult{\mbox{Mult}}
\def\iwi{\mbox{IW}}
\def\E{\text{E}}
\def\V{\text{Var}}

\def\D{\text{D}}

\def\P{\text{P}}

\def\d{\text{d}}

\def\bm{{\bf m}}
\def\bp{{\bf p}}

\def\bI{{\bf I}}
\def\bM{{\bf M}}

\def\bzero{{\bf 0}}
\def\simind{\stackrel{\mbox{\scriptsize{ind}}}{\sim}}
\def\simiid{\stackrel{\mbox{\scriptsize{iid}}}{\sim}}

\newcommand{\balpha}{\boldsymbol{\alpha}}

\newcommand{\bmu}{\boldsymbol{\mu}}

\newcommand{\bSigma}{\boldsymbol{\Sigma}}

\newcommand{\bPsi}{\boldsymbol{\Psi}}

\newcommand{\Ree}{{\rm I}\!{\rm R}}

\newcommand{\DP}{\mathcal{DP}}

\newcommand{\BB}{\mathcal{B}}

\newcommand{\MM}{\mathcal{M}}

\newcommand{\XX}{\mathcal{X}}

\begin{document}

\baselineskip=24pt

\title{{\bf Scalable Bayesian nonparametric measures for exploring pairwise dependence via Dirichlet Process Mixtures}}
\author{{\sc Sarah Filippi$^1$, Chris C. Holmes$^1$ and Luis E. Nieto-Barajas$^{1,2}$}\\[2mm]
{\sl $^1$Department of Statistics, University of Oxford, England}\\[2mm]
{\sl $^2$Department of Statistics, ITAM, Mexico}\\[2mm]
{\tt \small filippi@stats.ox.ac.uk, cholmes@stats.ox.ac.uk {\rm and} lnieto@itam.mx}}
\date{}

\maketitle

\begin{abstract}
In this article we propose novel Bayesian nonparametric methods using Dirichlet Process Mixture (DPM) models for detecting pairwise dependence between random variables while accounting for uncertainty in the form of the underlying distributions. A key criteria is that the procedures should  scale to large data sets. In this regard we find that the formal calculation of the Bayes factor for a dependent-vs.-independent DPM joint probability measure is not feasible computationally. To address this we present Bayesian diagnostic measures for characterising evidence against a  ``null model'' of pairwise independence.  In  simulation studies, as well as for a real data analysis, we show that our approach provides a useful  tool for the exploratory nonparametric Bayesian analysis of large multivariate data sets. 
\end{abstract}


\section{Introduction}
\label{sec:intro}

Identifying dependences among pairs of random variables measured on the same sample, producing datasets of the form $\D=\{(x_i,y_i),\;i=1,\ldots,n\}$, is an important task in modern exploratory data analysis where historically the Pearson correlation coefficient and the Spearman's rank correlation have been used. More recently there has been a move to the use of non-linear or distribution free methods such as those based on Mutual Information (MI) \citep{cover2012elements,kinney2014equitability}. In this paper we present Bayesian nonparametric methods for screening large data sets for possible pairwise associations (dependencies). Having an explicit probability measure of dependences has numerous advantages both in terms of interpretability and for integration  across different experimental conditions and/or  within a formal decision theoretic analysis. As data sets become ever larger and more complex we increasingly require Bayesian procedures that can scale to modern applications and this will be a key design criteria here. The main building block of our procedures will be the Dirichlet Process Mixture (DPM) model, which is the most popular Bayesian nonparametric model. 

We frame the problem of screening for evidence of pairwise dependence as a nonparametric model choice problem with alternatives: 
\begin{align}
\nonumber
\MM_0:& \text{ X and Y are independent random variables}\\
\label{eq:m0m1}
\MM_1:& \text{ X and Y are dependent random variables}\;.
\end{align}
Given a set of measurement pairs $\D$, for $n$ exchangeable observations one could then evaluate the posterior probability for competing models $\P(\MM_1 | \D) = 1 - \P(\MM_0 | \D)$ 
or consider the Bayes factor $\P(\D \mid \MM_0)/\P(\D \mid \MM_1)$ which is a measure of the strength of evidence for independence between the two samples against dependence. However with $p$ measurement variables under study there are $\approx \frac{1}{2}p^2$ such pairwise Bayes factors to compute, where even just one such evaluation might be problematic to compute. This motivates us to explore scalable alternatives to a formal Bayesian testing approach, by deriving summary statistics and functionals of the posterior that can provide strong indication in favour or against  independence.

Bayesian nonparametric hypotheses testing via Polya tree priors has been the focus of a couple of recent research papers \citep{holmes2015two,filippi2015bayesian}.
Here, however, we specify model uncertainty in the distribution of X and Y via DPMs of Gaussians. This provides flexibility while also encompassing smoothness assumptions on the underlying joint distributions. Another advantage is that DPMs have been widely studied in the Bayesian nonparametric literature with excellent open source implementation packages available \citep[e.g.][]{jara2011dppackage}. Moreover, although not explored here, the use of DPMs makes our approach readily extendable to situations when $X$ and $Y$ are themselves collections of multivariate measurements. Here we consider pairwise dependence between univariate measurements where for $\MM_0$, independence, the joint distribution factorises into a product of two univariate DPMs on $X$ and $Y$, while for $\MM_1$ we can define a joint DPM model on the bivariate measurement space $(X,Y)$.

In theory, given a DPM prior on the unknown densities, 
the Bayes factor can be calculated via the marginal likelihood. 
However this requires integrating over an infinite dimensional parameter space that does not have a tractable form. Moreover, using computational approaches to approximate the marginal likelihood is highly non-trivial, particularly when considering the need to scale to many thousands of comparisons with large $p$. 
To overcome this issue we present two new approaches to deriving scalable diagnostic measures corresponding to probabilistic measures of dependence, bypassing the need to calculate Bayes Factors that might not be feasible or desirable. Our methods are motivated by two recent proposals in the literature \citep{lock&dunson:13, kamary2014testing}, although neither of these papers consider the problem we address here as outlined below.

Our first approach utilises the well known latent allocation, or clustering, structure of the DPM model to induce a partition of the two-dimensional data space. By running a Gibbs sampler under the independence model the cluster allocation of observations to specific mixture components at each iteration can then be used to define a latent contingency table given by the mixture component memberships. For each of these contingency tables we  perform a parametric Bayesian independence-vs.-dependence test using conjugate multinomial-Dirichlet priors that lead to explicit analytic forms for the conditional marginal likelihoods. 
This proposal follows a similar idea considered in \cite{lock&dunson:13} who studied the two-sample testing problem. 
A key difference in what we present here, in addition to that we consider the problem of pairwise dependence, is that \cite{lock&dunson:13}  use a finite mixture model to induce a partition instead of an infinite nonparametric mixture model used here.

In our second approach, we adapt a recent procedure of \citep{kamary2014testing}, turning the model choice problem into an estimation problem by writing the competing models under a hierarchy that incorporates both models, $\MM^* = \pi \MM_1 + (1-\pi) \MM_0$. We investigate the specification of $\MM^*$ either as a mixture model with mixing component $0 \le \pi \le 1$, or as a predictive linear ensemble of the two sub-models with constraints on the weights. We then estimate $\pi$ which becomes a measure of the evidence for dependence. DPMs are used to obtain the likelihood associated to each of the competing models in $\MM^*$, requiring a separate MCMC run for each potential pair of random variables. 

We compare and contrast the two procedures with particular regard to their scalability to large data sets. This latter feature naturally includes the amenity of the methods to simulation with modern parallel computation. We demonstrate that our association measures are scalable and successfully detect some highly non-linear dependences with equivalent performance to the current best conventional methods using mutual information, with the added advantages that fully probabilistic Bayesian methods enjoy. As mentioned above, some of these key advantages includes the ability to integrate results within a formal decision analysis framework, or within optimal experimental design, and the combination of results with other sources of information, or across studies such as arise in meta-analysis.

The rest of the paper is as follows. In Section~\ref{sec:DPM} we review the Dirichlet Process and the DPM of Gaussians. In Section~\ref{sec:method} we describe the two approaches to quantify the evidence for dependence using Dirichlet Process Mixtures. In Section~\ref{sec:numerical} we illustrate our approach on the exploratory analysis of a real-world example from the World Health Organisation data set of country statistics and also on simulated data generated from simple models. We conclude the paper with a short discussion in Section~\ref{sec:conclusion}.

\section{Dirichlet Process Mixtures}
\label{sec:DPM}

The Dirichlet process \citep{ferguson:73} is the most important process prior in Bayesian nonparametric statistics. It is flexible enough to approximate (in the sense of weak convergence) any probability law, although the paths of the process are almost surely discrete \citep{blackwell&macqueen:73}. Many years ago this discreteness was considered a drawback but nowadays it is simply a feature that characterises the Dirichlet process. This feature has recently been highly exploited in clustering applications (e.g. \citep{dahl:06}).

The Dirichlet process is defined as follows. Let $G$ be a probability
measure defined on $(\XX,\BB)$, where $\XX\subset\Ree^p$ and $\BB$ the
corresponding Borel's $\sigma$-algebra. Let $G$ be a stochastic process
indexed by the elements of $\BB$. $G$ is a Dirichlet process with
parameters $c$ and $G_0$ if for every measurable partition
$(B_1,\ldots,B_k)$ of $\XX$,
$$(G(B_1),\ldots,G(B_k))\sim\dir(cG_0(B_1),\ldots,cG_0(B_k)).$$
From here we can see that, for every $B\in\BB$, $\E\{G(B)\}=G_0(B)$ and
$\V\{G(B)\}=G_0(B)\{1-G_0(B)\}/(c+1)$. Therefore the parameter $c$ is
known as precision parameter and $G_0$ as the centering measure.

The Dirichlet process when used as a priori induces exchangeability in the data. In notation, let $X_1,\ldots,X_n$
be a sample of random variables such that conditional on $G$, $X_i\mid
G\simiid G$. If we further take $G\sim\DP(c,G_0)$ then the marginal
distribution of the data $(X_1,\ldots,X_n)$ once the process $G$ has been
integrated out, is characterised by what is known as the P\'olya urn
\citep{blackwell&macqueen:73}. We start with $X_1\sim G_0$ then
\begin{equation}
\label{eq:polyaurn}
X_{n}\mid X_1,\ldots,X_{n-1} \sim
\frac{cG_0+\sum_{j=1}^{n-1}\delta_{X_j}}{c+n-1}.
\end{equation}

Instead of placing the Dirichlet process prior directly on the observable
data, it can be used as the law of the parameters of another model
(kernel) that generated the data. In notation, let us assume that for each
$i=1,\ldots,n$, $$X_i\mid\theta_i\simind f(x_i\mid\theta_i),$$ with $f$ a
parametric density function. We can further take $$\theta_i\mid G \simiid
G$$ with $$G\sim\DP(c,G_0).$$ This hierarchical specification can be seen
as a mixture of density kernels $f(x\mid\theta)$ with mixing distribution
coming from a Dirichlet process, i.e., $\int f(x\mid\theta)G(\d\theta)$.
This model is known as Dirichlet process mixture (DPM) model and was first
introduced by \cite{lo:84} in the context of density estimation and
written in hierarchical form by \cite{ferguson:83}.

The most typical choice of kernel $f$ is the (multivariate) normal, in
which case $\theta_i=(\mu_i,\sigma_i^2)$, with scalars mean and variance, in the univariate case, and
$\theta_i=(\bmu_i,\bSigma_i)$, with mean vector and variance-covariance matrix, in the multivariate case. We will work with this specific kernel throughout this paper.

As can be seen by construction \eqref{eq:polyaurn}, in the mixture case,
the Dirichlet process induces a joint distribution on the set
$(\theta_1,\ldots,\theta_n)$ that allows for ties in the $\theta_i$'s.
This in turn induces a clustering structure in the $\theta_i$'s (and
$X_i$'s). Posterior inference of the DPM model usually relies on a Gibbs
sampler \citep{smith&roberts:93}. At each iteration of the Gibbs sampler
the model produces a different clustering structure. The number of
clusters is a function of the sample size $n$ and the precision parameter
$c$ of the underlying Dirichlet process. The larger the value of $c$, the
larger the number of clusters induced. This clustering structure and parameter $c$ will play a central
role in one of the independence test procedures that will be described later.

\section{Two approaches for measuring dependence}
\label{sec:method}

As noted in Section 1, the calculation or approximation of the formal Bayes factor under $\MM_0$ and $\MM_1$ is not feasible when considering a large number of model comparisons. Indeed it may not even be desirable given that our objective is to highlight potential departures from independence rather than answer a formal model choice question.  
In this section we describe two distinct approaches for comparing models $\MM_0$ and $\MM_1$ defined in \eqref{eq:m0m1} based on DPM models that are computable and scalable to large data. 

\subsection{Contingency tables approach}

The first approach is motivated by the paper from \cite{lock&dunson:13} who turned a two-sample testing problem into a discrete test on the clustered data. Recall that the two-sample testing problem considers the same measurement variable recorded on separate subjects under two different conditions; whereas we are considering different measurement variables recorded on the same subject. Similar to  \cite{lock&dunson:13}, our procedure consists in marginally discretising the data into ordered categories and performing a Dirichlet-multinomial independence test on the induced contingency table. This amounts to first clustering the data under $\MM_0$ and then exploring for evidence of departure from $\MM_0$, toward $\MM_1$, by testing for statistical association between the cluster memberships in $X$ and $Y$. Uncertainty in the cluster memberships is accounted for by the DPM defined under $\MM_0$, as outlined below.

To begin assume that the data are marginally clustered in $K_X$ and $K_Y$ clusters and denote by $\xi_{X,i}\in \{1\ldots,K_X\}$ and 
$\xi_{Y,i}\in \{1,\dots,K_Y\}$ the cluster indicators for the data points $x_i$ and $y_i$ respectively, for $i=1,\dots,n$.  Using these cluster indicators, we can construct a contingency table $\bM_{\xi_X,\xi_Y} =\{m_{kl}\}$ of size $K_X\times K_Y$, such that $m_{kl}=\sum_{i=1}^n I(\xi_{X,i}=k,\, \xi_{Y,i}=l)$, for $k=1,\dots K_X$ and $l=1,\dots,K_Y$. 
The contingency table $\bM_{\xi_X,\xi_Y}$ represents a discretised version of the (unnormalised) marginals and joint distribution of the continuous vector $(X,Y)$. We can then apply Bayesian independence tests for discrete / categorical variables following \cite{gunel&dickey:74} and \cite{good&crook:87} who proposed a conjugate multinomial-Dirichlet independence test which is described as follows. 
Let $\bM_{\xi_X,\xi_Y}\sim\mult(n,\bp)$ with $\bp=\{p_{kl}\}$ the matrix of cell probabilities of dimension $K_X\times K_Y$.
Consider a 
conjugate prior distribution $\bp\sim\dir(\balpha)$, with $\balpha=\{\alpha_{kl}\}$ such that $\sum_{kl}\alpha_{kl}=a$.  In practice we suggest to use $\alpha_{kl}=a ( K_XK_Y)^{-1}$ or $\alpha_{kl}=1/2$ for all $1\leq k\leq K_X$ and $1\leq l\leq K_Y$. Under model $\MM_1$ the probability of having observed the counts in $\bM_{\xi_X,\xi_Y}$ 
is
\begin{equation}
\label{eq:pm0}
\P(\bM_{\xi_X,\xi_Y}\mid\MM_1, \xi_{X}, \xi_{Y})=\int \P(\bM_{\xi_X,\xi_Y}\mid\bp) f(\bp)\,\d\bp
=\frac{\Gamma(a)}{\Gamma(a+n)}\prod_{k,l}\frac{\Gamma(\alpha_{kl}+m_{kl})}{\Gamma(\alpha_{kl})}.
\end{equation}
Under the independent model $\MM_0$ the observed counts $\bM_{\xi_X,\xi_Y}$ can be expressed in terms of the marginal counts $\bm_X=\{m_{k\cdot}\}$ and $\bm_Y=\{m_{\cdot l}\}$ whose implied distributions are again multinomial with probability vectors $\bp_X=\{p_{k\cdot}\}$ and $\bp_Y=\{p_{\cdot l}\}$, respectively, with $p_{k\cdot}=\sum_{l}p_{kl}$ and $p_{\cdot l}=\sum_{k}p_{kl}$. The induced prior distributions are also Dirichlet with parameters $\balpha_X=\{\alpha_{k\cdot}\}$ and $\balpha_Y=\{\alpha_{\cdot l}\}$. Then, the probability of $\bM_{\xi_X,\xi_Y}$ under $\MM_0$ becomes 
\begin{align}
\nonumber
\P(\bM_{\xi_X,\xi_Y}\mid\MM_0, \xi_{X}, \xi_{Y})&=\int \P(\bm_X\mid\bp_X)f(\bp_X)\,\d\bp_X \int \P(\bm_Y\mid\bp_Y)f(\bp_Y)\,\d\bp_Y \\
\label{eq:pm1}
&=\frac{\Gamma^2(a)}{\Gamma^2(a+n)}\prod_{k}\frac{\Gamma(\alpha_{k\cdot}+m_{k\cdot})}{\Gamma(\alpha_{k\cdot})}\prod_{l}\frac{\Gamma(\alpha_{\cdot l}+m_{\cdot l})}{\Gamma(\alpha_{\cdot l})}\;,
\end{align} 
where $\alpha_{k\cdot}=\sum_{l}\alpha_{kl}$ and $\alpha_{\cdot l}=\sum_{k}\alpha_{kl}$. 

To compare evidence in favour of each model, we use expressions \eqref{eq:pm0} and \eqref{eq:pm1} to compute the Bayes factor $BF_{\xi}=\P(\bM_{\xi_X,\xi_Y}\mid\MM_0, \xi_{X}, \xi_{Y})/\P(\bM_{\xi_X,\xi_Y}\mid\MM_1, \xi_{X}, \xi_{Y})$. 
Using equal prior probabilities for both models, i.e. $\P(\MM_0)=\P(\MM_1)=0.5$, we obtain that the posterior probabilities for the independence and dependence models are 
$\P(\MM_1\mid\bM_{\xi_X,\xi_Y})=1/(1+BF_{\xi_X,\xi_Y})=1-\P(\MM_0\mid\bM_{\xi_X,\xi_Y}).$ where 
\begin{equation}
BF_{\xi_X,\xi_Y}=\frac{\Gamma(a)}{\Gamma(a+n)}\prod_{k}\frac{\Gamma(\alpha_{k\cdot}+m_{k\cdot})}{\Gamma(\alpha_{k\cdot})}\prod_{l}\frac{\Gamma(\alpha_{\cdot l}+m_{\cdot l})}{\Gamma(\alpha_{\cdot l})}\prod_{k,l}\frac{\Gamma(\alpha_{kl})}{\Gamma(\alpha_{kl}+m_{kl})}\;.
\label{eq:BFinCT-BF}
\end{equation}
It should also be noted that this contingency table approach would also afford a conditional frequentist test. For example, consider 
Pearson's chi-squared test of independence \citep{pearson:22}. Denote by 
$m_{k\cdot}=\sum_l m_{kl}$ and $m_{\cdot l}=\sum_k m_{kl}$ the number of individuals classified in cluster $k$ of $X$ and cluster $l$ of $Y$, respectively. Then, the well known test statistic is 
\begin{equation}
\label{eq:CTpvalue}
T=\sum_{k=1}^{K_X}\sum_{l=1}^{K_Y}\frac{(m_{kl}-m_{k\cdot}m_{\cdot l}/n)^2}{m_{k\cdot}m_{\cdot l}/n}\,.
\end{equation}
Under the null hypothesis $\MM_0$ of independence, statistic $T$ follows a $\chi^2$ distribution with $(K_X-1)(K_Y-1)$ degrees of freedom. If the test statistic is improbably large according to that chi-square distribution, then one rejects the null hypothesis $\MM_0$ in favour of the dependence hypothesis $\MM_1$.

The hypothesis testing approach described in this section assumes that the data are marginally clustered. However, these clusters are not known a prior. A Bayesian approach for data clustering is to define a prior distribution over the clustering and then update the posterior based on the evidence provided by the data. Here we make use of the DPM model structure to create an empirical partition of the two-dimensional data space, taking into account the uncertainty on the allocation process. More precisely,
we consider two independent DPM prior models for each of the marginal densities with the following specifications:  \begin{equation}
\label{eq:prior0a}
f_{0,X}(x)\sim\int\no(x\mid\theta_X)G_X(\d\theta_X)\;\;\;\text{and}\;\;\; f_{0,Y}(y)\sim\int\no(y\mid\theta_Y)G_Y(\d\theta_Y),
\end{equation}
where $\theta_X=(\mu_X,\sigma_X^2)$ and $\theta_Y=(\mu_Y,\sigma_Y^2)$, with 
\begin{equation}
\label{eq:prior0b}
G_X\sim\DP(c_0,G_0)\;\; \text{and}\;\; G_Y\sim\DP(c_0,G_0)
\end{equation} 
and $G_0=\no(\mu\mid \mu_0,\sigma^2/k_0)\,\iga(\sigma^2\mid \nu/2-1,\psi/2)$. 
The latent clustering structure induced by the DPM models defined by \eqref{eq:prior0a} and \eqref{eq:prior0b}, can then be used to construct a contingency table as described above. Note that in an ideal world one would carefully specify subjective beliefs on the prior marginals for $X$ and $Y$. However, when the number of variables is large this is not feasible and we require some default specification as done here, by assuming a common prior after suitable transformation of the data. 

 Although it is clear from the properties of the DP that it induces a partition, in practice it is not easy to determine an optimal one. 
Fitting a DPM model via a Gibbs sampler provides a partition at each iteration. We can proceed in two different ways. One is to use all potential partitions coming from the MCMC, and for each of them perform the Bayesian independence test and report the expected posterior probability. More precisely, the functional we consider is 
\begin{equation}
p_\text{dep}=\int \frac{1}{(1+BF_{\xi_X,\xi_Y})}p(\xi_{X}, \xi_{Y})d\xi_Xd\xi_Y\;. \label{eq:pindepCTBF}
\end{equation}

This is the procedure we recommend and develop below. An alternative approach would be to consider the selection of one of the partitions using an appropriate optimization criterion, for example using the criterion of \cite{dahl:06} who proposes to choose the partition that minimises the squared deviations with respect to the average pairwise clustering matrix, and use that single partition to perform the test,  ignoring the uncertainty in the partition structure as in \cite{lock&dunson:13} for the two-sample test. In \textit{Supplementary Material} we provide an empirical comparison between both procedures.

In the rest of the paper we will focus on the first alternative that considers all potential partitions; we will refer to this procedure as CT-BF. 



\begin{algorithm}                      
\caption{Independence measure based on Contingency table (CT-BF)}          
\label{alg:CT}                           
\begin{algorithmic}                    
   \REQUIRE Data $\D=\{x_i,y_i\}_{i=1}^n$
    \REQUIRE Prior parameters $a$
    \REQUIRE Prior parameters for the DPM and number of iterations $N_\text{it}$
    \ENSURE Probability of dependence $p_\text{dep}$     \vspace{0.2cm}
    \STATE \underline{DPM inference:}
    \STATE Infer a DPM model for the distribution $f_{0,X}(x)$ using a Gibbs Sampler with $n_\text{it}$ iterations
    \STATE $\rightarrow$ for each iteration $1\leq j\leq N_\text{it}$, record a vector of cluster indicator $\xi_X^{(j)}$ 
    \STATE  Infer a DPM model for the distribution $f_{0,Y}(y)$ using a Gibbs Sampler with $N_\text{it}$ iterations
    \STATE $\rightarrow$ for each iteration $1\leq j\leq N_\text{it}$, record a vector of cluster indicator $\xi_Y^{(j)}$ 
   \vspace{0.2cm}
    \FOR{$1\leq j\leq N_\text{it}$}
    \STATE Construct a contingency table $\bM^{(j)}$ of size $K_X^{(j)}\times K_Y^{(j)}$ based on $\xi_X^{(j)}$  and $\xi_Y^{(j)}$ 
    \STATE $p^{(j)} \leftarrow 1/(1+BF)$
    where $BF$ is defined in \eqref{eq:BFinCT-BF}.
    \ENDFOR
     \vspace{0.2cm}
    \STATE $p_\text{dep}\leftarrow\frac{1}{n_\text{it}}\sum_{j=1}^{n_\text{it}}p^{(j)}$ 
\end{algorithmic}
\end{algorithm}

\subsection{Mixture model predictive approach}

In this section we consider an alternative approach for testing between hypothesis \eqref{eq:m0m1}. Motivated by \cite{kamary2014testing} we replace the testing problem with an estimation one by defining a predictive ensemble model $\MM^*$ whose components are the competing models $\MM_0$ and $\MM_1$. 
To be precise, let $f_0$ and $f_1$ denote the densities of $(X,Y)$ defined by models $\MM_0$ and $\MM_1$, respectively. Then we define a  predictive mixture model as a linear combination of sub-models of the form
\begin{equation}
\label{eq:mixmod}
f^*(x,y) = \pi f_1(x,y)+(1-\pi) f_0(x,y),
\end{equation}
where 
 $\pi$ is a free regression parameter with constraint $0\leq \pi\leq 1$ and $ f_0(x,y)=f_{0,X}(x)f_{0,Y}(y)$. This model embeds both $\MM_0$ and $\MM_1$ for values of $\pi$ equal to $0$ or $1$. The main idea of this method is to estimate from the data the mixture parameter $\pi$, which indicates the preference of the data for dependence model $\MM_1$. In contrast to the latent contingency table procedure this approach requires the explicit construction of a joint model under hypothesis ${\cal{M}}_1$.

Since $f_0$ and $f_1$ are unknown densities, we assume Bayesian nonparametric prior distributions. 
For $f_{0_X}(x)$ and $f_{0,Y}(y)$ we consider the DPM model defined by equations \eqref{eq:prior0a} and \eqref{eq:prior0b}. For $f_1$ we take a bivariate DPM model defined as
\begin{equation}
\label{eq:prior1a}
f_1(x,y)\sim\int\no(x,y\mid\theta_{X,Y})G_{X,Y}(\d\theta_{X,Y}),
\end{equation} 
where $\theta_{X,Y}=(\bmu,\bSigma)$, with 
\begin{equation}
\label{eq:prior1b}
G_{X,Y}\sim\DP(c_{1},G_{1})
\end{equation} 
and $G_{1}=\no(\bmu\mid \bmu_0,(1/k_0)\bSigma)\,\iwi(\bSigma\mid\nu,\bPsi)$.
The parameter $\pi$ has also to be estimated so we take a prior of the form $\pi\sim\be(a_0,b_0)$.  We ensure that the centring measures $G_0$ and $G_1$ are comparable by setting their hyper-parameters  as follows: we have $G_{d-1}=\no(\bmu\mid \bmu_0,(1/k_0)\bSigma)\,\iwi(\bSigma\mid\nu,\bPsi)$ for $d=1$ and $2$ with  $\nu=d+2$, the $d$-dimensional vector $\bmu_0\sim\no(0_d,c_\mu\;\bI_d)$, the $d\times d$-matrix $\bPsi\sim\iwi(\nu,c_\Psi \;\bI_d)$ where $\bI_d$ is the identity matrix of dimension $d$. The hyper-parameters $c_\mu$, $c_\Psi$ and $k_0$ are set to be equal for $G_0$ and $G_1$.

Our objective is to highlight pairwise dependence across many pairs of variables, and order the pairs into those showing evidence from strongest to weakest association. This motivates us to consider a simplified method by assessing the relative posterior predictive evidence under $\MM_0$ to that of $\MM_1$, by calculating an ensemble model using the posterior predictive probability of the observed data $f_1(x_{new},y_{new}|D)$ and $f_0(x_{new},y_{new}|D)$ separately. In the following we will use the notations $\hat{f}_j(x_{new},y_{new})=f_j(x_{new},y_{new}|D)$, $j=0,1$ to denote the posterior predictive distribution. It is important to note that for all $[p \times (p-1) / 2]$ $X,Y$ pairs we use the same prior, and hence same model complexity across all pairs, so ranking by the improvement in posterior predictive likelihood under $\MM_1$ relative to $\MM_0$ should not {\em{a priori}} favour certain pairs over others. This procedure significantly simplifies the inference as we can infer the posterior models by first fitting the three DPM models separately each using the entire sample data, and then updating the ensemble parameter $\pi$ from its posterior conditional distribution $$f(\pi\mid\D)\propto f(\pi)\prod_i \left(\pi \hat{f}_1(x_i,y_i)+(1-\pi)\hat{f}_0(x_i,y_i)\right)\;,$$which is a simple line search on $[0,1]$. We will refer to this inference procedure as MixMod-ensemble -- see Algorithm \ref{alg:MixMod2steps}.

\begin{algorithm}                      
\caption{Independence test MixMod-ensemble }          
\label{alg:MixMod2steps}                           
\begin{algorithmic}                    
 \REQUIRE Data $\D=\{x_i,y_i\}_{i=1}^n$; Prior parameters $a_0$ and $b_0$; Prior parameters for the DPMs
    \ENSURE Estimate of mixture parameter $\pi$
    \vspace{0.2cm}
    \STATE \underline{DPMs inference:}
        \STATE $\hat{f}_{0,X} \leftarrow$ posterior prediction of a DPM for distribution of $\{x_i\}_i$ averaged over all Gibbs sampler iteration
         \STATE $\hat{f}_ {0,Y}\leftarrow$ posterior prediction of a DPM for distribution of $\{y_i\}_i$ averaged over all Gibbs sampler iteration
          \STATE $\hat{f}_{1} \leftarrow$ posterior prediction of a DPM for distribution of $\{x_i,y_i\}_i$ averaged over all Gibbs sampler iteration
      \vspace{0.2cm}
      \STATE \underline{Estimation of $\hat{\pi}$:}
      \STATE Define a fine grid of $[0,1]$ with intervals of length $\eta=10^{-4}$
    \FOR{$j=0, \dots, \eta^{-1}$}
    \STATE $\pi^{(j)}\leftarrow j\times \eta$
    \STATE $L_j \leftarrow \sum_{i=1}^n\log(\pi^{(j)}\hat{f}_1(x_i,y_i)+(1-\pi^{(j)})\hat{f}_{0,X}(x_i)\hat{f}_{0,Y}(y_i))+\log(\be(\pi^{(j)}\mid a_0,b_0))$
    \ENDFOR
    \STATE $\hat{\pi}\leftarrow \frac{1}{\sum_j\exp(L_j)}\sum_j\pi^{(j)}\exp(L_j)$
\end{algorithmic}
\end{algorithm}

An alternative approach, more closely resembling \cite{kamary2014testing}, is to consider $\MM^*$  as a mixture-model rather than an ensemble model where with probability $\pi$ the data arises from $f_0$ and with probability $1-\pi$ from $f_1$. \cite{diebolt&robert:94} show that posterior sampling in a mixture model is simplified if we introduce latent variable indicators $\zeta_i\sim\ber(\pi)$ that determine whether observation $i$ comes from $f_1$, when $\zeta_i=1$, or from $f_0$, when $\zeta_i=0$. Conditional on these latent indicators the mixture components $f_0$ and $f_1$ can be updated using only the data points allocated to each model. As noted by \cite{kamary2014testing}, the Gibbs sampler implemented in this way can become quite inefficient if the parameter $\pi$ approaches the boundaries $\{0,1\}$, specially for large sample sizes.  We refer to this method as MixMod. For our purposes this  requires specifying a Gibbs sampler for the mixture model utilising three DPM models $\{f_1(x,y), f_{0,X}(x), f_{0,Y}(y)\}$ and the mixture allocations for points across all $p \times (p-1) / 2$ pairs.

In the paper we will illustrate the performance using MixMod-ensemble, and in the \textit{Supplementary Material} we provide a comparison between MixMod and MixMod-ensemble.

Regardless of the posterior inference procedure, different estimators of $\pi$ could be obtained from its posterior distribution. We chose to select the expected value as a statistic of dependence, that is, 
\begin{equation}
\hat{\pi}=\E(\pi\mid\D)=\int_0^1\pi\; f(\pi \mid D)d\pi\;.
\label{eq:MM}
\end{equation}

\subsection{Computational tractability}
Both of the Bayesian non-parametric approaches proposed here are motivated by the increasing necessity of screening large data sets for possible pairwise dependencies where calculation of the formal Bayes factor under $\MM_0$ and $\MM_1$ is unfeasible or undesirable. In this section, we discuss some computational advantages of our two methods including their amenity to implementation on modern computing architectures exploiting parallelisation on multi-core standalone machines, or clusters of multi-core and many-core machines, or cloud based computing environments.

In relation to parallelisation we see that both methods are divided in two steps: one starts by inferring DPMs using a Gibbs sampler and then perform a dependence test using every iteration of the Gibbs sampler. This decoupling of the inference step and the model comparison step allows to significantly reduce the computational cost of the procedure. In particular, only a couple of thousands of Gibbs sampling iterations are necessary to estimate the predictive posterior densities and posterior distributions over the latent allocation variables. In the environment for statistical computing \cite{rpackage}, the parallelisation of both approaches is very simple and only consists in replacing the command \textit{apply} by the command \textit{parLapply} from the package \textit{parallel}  -- which is included in versions of R following 2.14.0. The R code to run CT-BF and MixMod-ensemble independence tests is available in the \textit{Supplementary Material}.

The CT-BF approach based on the construction of a contingency table is particularly attractive as it is trivially parallelizable and does not involve an explicit DPM model for the joint $f_1(x,y)$ under $\MM_1$. With $p$ measurement variables under study, this approach only needs to infer $p$ independent marginal DPMs, recording information from $N_\text{it}$ Gibbs sampling iterations for each of them independently in parallel. The MCMC output from the $p$ models is then combined and we perform $N_\text{it} \times p\times(p-1)/2$ independent tests where following (\ref{eq:BFinCT-BF}) only involves computing ratios of Gamma functions. As an illustration, in the example described in more details in Section 4, for $p=562$ measurement variables, the first stage of inference on the DPMs take less than $3$ minutes on a $48$-core machine, and then the resulting $1.5\times 10^8$ pairwise tests of dependence for all pairs of variables are performed in one hour. 

In comparison the MixMod-ensemble approach incurs a greater computational overhead as we require bivariate DPMs, $f_1(x,y)$, to be fit for all pairs. In the illustration below the MixMod-ensemble procedure for the $1.5\times 10^8$ pairs takes approximatively 36 hours on the same 48-core machine.

\section{Numerical Analysis}
\label{sec:numerical}

\subsection{World Health Organisation dataset}
\label{sec:WHO}
\begin{figure}
\includegraphics[width=\textwidth]{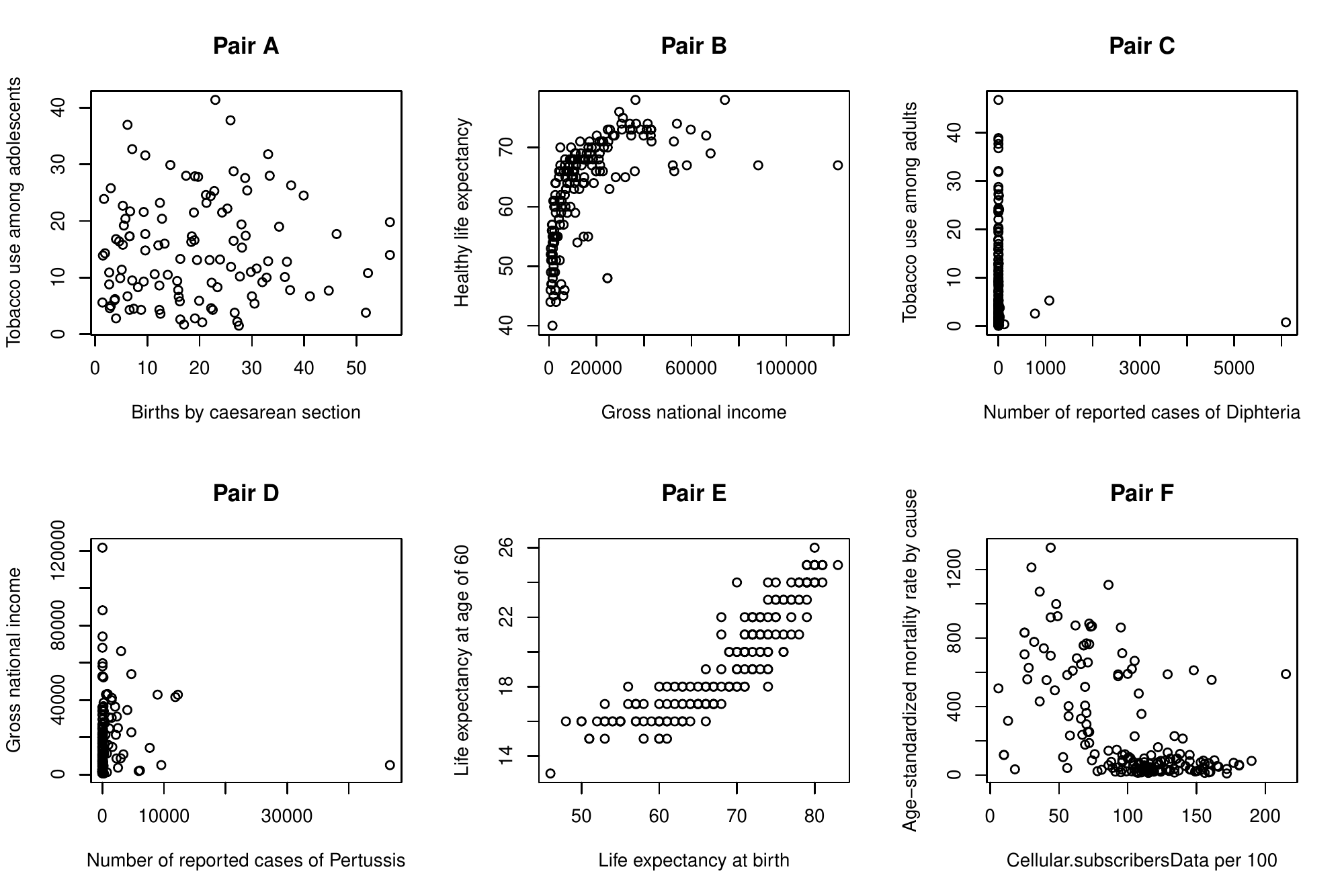}
\caption{Examples of the relationship between economic, social and health indicators provided by the WHO Statistical Information System. Each dot corresponds to one country.}
\label{fig:whoData}
\end{figure}

In this section, we apply the two approaches described in Section \ref{sec:method} to detect dependencies in economic, social and health indicators from the World Health Organisation (WHO). The WHO Statistical Information System (WHOSIS) has recently been incorporated into the Global Health Observatory (GHO) that contains a data repository (\url{http://www.who.int/gho/database/en/}) with mortality and global health estimates, demographic and socioeconomic statistics as well as information regarding health service coverage and risk factors for $194$ countries. We combined these datasets to obtain a set of $562$ statistics per country. We aim at highlighting potential dependencies between these indicators. Scatterplots of some of these indicators are represented in Figure~\ref{fig:whoData}, where for example we see, unsurprisingly, strong dependencies between indicators such as life expectancy at birth and increased life expectancy at age 60 (Pair E). 

We applied both the CT-BF and the MixMod-ensemble test to compute the probability of dependence for all the 157,641 pairs of indicators.
The two proposed methods require the specification of several parameters of the prior distributions. The impact of these choices is discussed in \textit{Supplementary Material}.  For the approach based on contingency tables the prior specifications for models \eqref{eq:prior0a} and \eqref{eq:prior0b} are set as follows: $c_0=10$, $\mu_0\sim\no(0,1)$, $k_0\sim\ga(1/2,100/2)$, $\nu=3$ and $\psi\sim\iga(1/2,5)$. Note that $c_0$ controls the number of clusters induced, so in order to avoid having partitions with only one cluster we set this parameter at a relative large value. To specify the Dirichlet prior for the cell probabilities in the contingency table we took $\alpha_{kl}=1/2$, which is the Jeffreys prior in a multinomial model. In experimentation we found that the contingency table can be sensitive to the choice of the parameter $c_0$. This parameter influences the number of clusters in the DPM model and therefore the size of the contingency tables and it is important to specify a value that induces a reasonable number of clusters. We would recommend exploring several values. Results seem fairly insensitive to the choice of the parameters $\alpha_{kl}$ in the Dirichlet priors. 

For the approach considering an ensemble mixture model, 
the parameters $c_0$ and $c_1$ are not fixed but specified by $c_0, c_1\sim\ga(1,1)$ and $\mu_0\sim\no(0,100)$. This change was introduced to allow the model to determine the best fit without constraining the number of clusters. In addition, the prior processes $G_0$ and $G_1$ are defined as follows: $G_{d-1}=\no(\bmu\mid \bmu_0,(1/k_0)\bSigma)\,\iwi(\bSigma\mid\nu,\bPsi)$ for $d=1$ and $2$ with  $\nu=d+2$, the $d$-dimensional vector $\bmu_0\sim\no(0_d,100\;\bI_d)$, the $d\times d$-matrix $\bPsi\sim\iwi(\nu, 0.1\;\bI_d)$ and $k_0\sim\ga(1/2,50)$, where $\bI_d$ is the identity matrix of dimension $d$. The prior distribution of the mixing proportion $\pi$ was specified by taking $a_0=b_0=1/2$. Our experience is that results are fairly robust to the prior parameter settings (see \textit{Supplementary Material}).
 
The procedures were implemented in the environment for statistical computing \cite{rpackage} and make use of the package \emph{DPpackage} \citep{jara2011dppackage}. Chains were run for 10,000 iterations with a burn in of 1,000 keeping one of every 5th draws for computing estimates. 

For both approaches the tests were performed only for pairs containing measurements for at least 10 countries. 
For the CT-BF approach, the $562$ DPMs are inferred using all the available data; however, the contingency tables were constructed taking into account only the countries for which both indicators (in the pair) are available. 
For the MixMod-ensemble approach, in order to avoid any bias towards one of the two models $\MM_0$ or $\MM_1$, both the DPMs on the marginals and the DPM on the joint space are inferred only on the countries for which measurements are available for both indicators. Extending the method to handle missing data is a future objective.

\begin{figure}
\includegraphics[width=\textwidth]{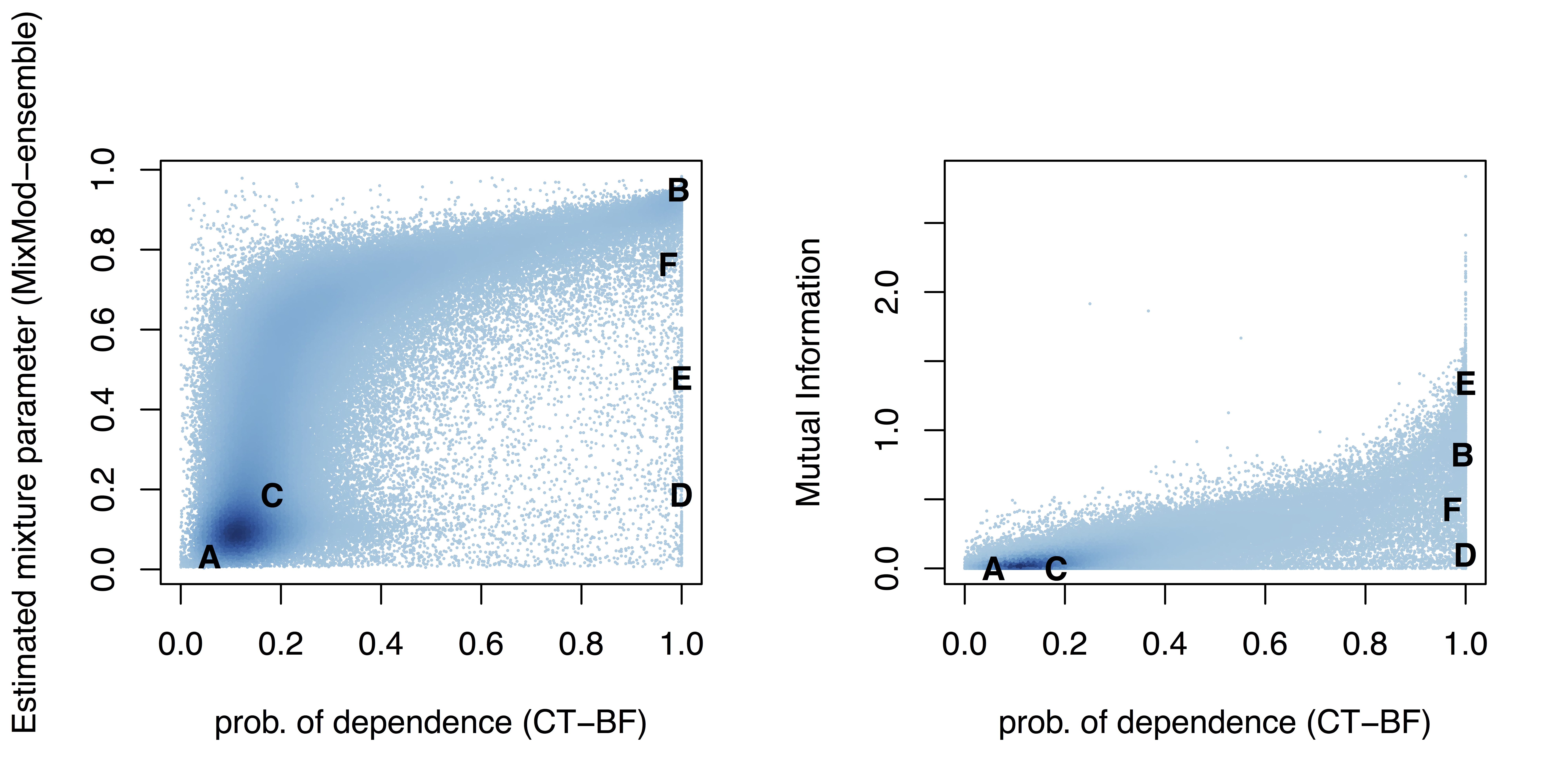}
\caption{Performance comparison between the CT-BF and the MixMod-ensemble approaches (left) and the mutual information (right) for every pair of indicators in the WHO dataset. The probabilities  of dependences obtained following CT-BF and MixMod-ensemble are respectively $p_\text{dep}$ and  $\hat\pi$, defined equations~\eqref{eq:pindepCTBF} and \eqref{eq:MM} and approximated following algorithms \ref{alg:CT} and \ref{alg:MixMod2steps}.  The letters A to F correspond to the 6 pairs of indicators illustrated in Figure~\ref{fig:whoData}.}
\label{fig:whoResults}
\end{figure}

The measure of dependences obtained following our two approaches, i.e. $p_\text{dep}$ for CT-BF and $\hat\pi$ for MixMod-ensemble, defined respectively equations~\eqref{eq:pindepCTBF} and \eqref{eq:MM},  are compared for each pair of variables in Figure~\ref{fig:whoResults} (left panel). Strong dependences (defined as $p_\text{dep}> 0.8$) are detected for 5\% of pairs, and credible independence (i.e. $p_\text{dep}< 0.2$) between 30\% of the indicators. We observe that the two probabilistic measures of dependence generally agree for most of the pairs, with the probability value obtained following the MixMod-ensemble method being generally higher than the probability measure obtained following the CT-BF approach. This elevation in the evidence in dependence is perhaps to be expected as MixMod-ensemble uses the conditional posterior predictive likelihood which will favour the more complex joint model of $f_1(x,y)$. However, the two methods disagree (defined as the probability value obtained following one method is lower than $0.2$ while it is larger than $0.8$ following the other method) for less than 0.36\% of the pairs; and these differences mainly occur when one of the $(X,Y)$ variables is equal to $0$ for more than 20\% of the countries (see for example pairs C and D). 

On balance we prefer to use the CT-BF approach due to its computational scalability, 1 hour of run-time on a 48-core computer in comparison with 36 hours for MixMod-ensemble in this example. 
We compared the analysis from the CT-BF to that using a mutual information approach computed using the 20-nearest neighbours method, as in \cite{kinney2014equitability} (see Figure~\ref{fig:whoResults} right panel where the labelled points correspond to plots in Fig~\ref{fig:whoData}).  We remark that some pairs of variables with strong dependences under CT-BF have a wide spread of   mutual information, in particular we note pairs D and F that have a probability of dependence close to 1 for CT-BF but relatively low MI values. Visually at least one could argue that associations of the form seen in Figure~\ref{fig:whoResults} D and F may be of potential interest to follow up by the analyst. 

\subsection{Simulation Study for frequentist power analysis}
\label{sec:simulated}
In this section we perform a simulation study to examine the frequentist performance of the two proposed tests on some controlled scenarios. The objective is to verify that we are not losing much power against a popular non-probabilistic method based on mutual information, which is optimised for frequentist power. Simulated datasets are generated under the following four different scenarios: 
\begin{enumerate}
\item A bivariate normal model: $(X,Y)\sim\no_2(\bzero,\bSigma)$ with $\bSigma=\left(\begin{array}{c c} 1 & \rho\\ \rho & 1 \end{array}\right)$, 
\item A sinusoidal model: $Y = 2\sin(X) + \eta$, with $\eta\sim\no(0,\phi^2)$, and $X\sim \un [0,5\pi]$
\item A parabolic model: $Y=2X^2/3 +\eta$, with $\eta\sim\no(0,\phi^2)$, and $X\sim\no(0,1)$
\item A circular model: $X = 10\cos(\theta) + \eta$ and $Y = 10\sin(\theta) + \eta$, with $\theta\sim\un[0,2\pi]$ and $\eta\sim\no(0,\phi^2)$. 
\end{enumerate}

\begin{figure}[H]
\centering
\includegraphics[width=0.8\textwidth]{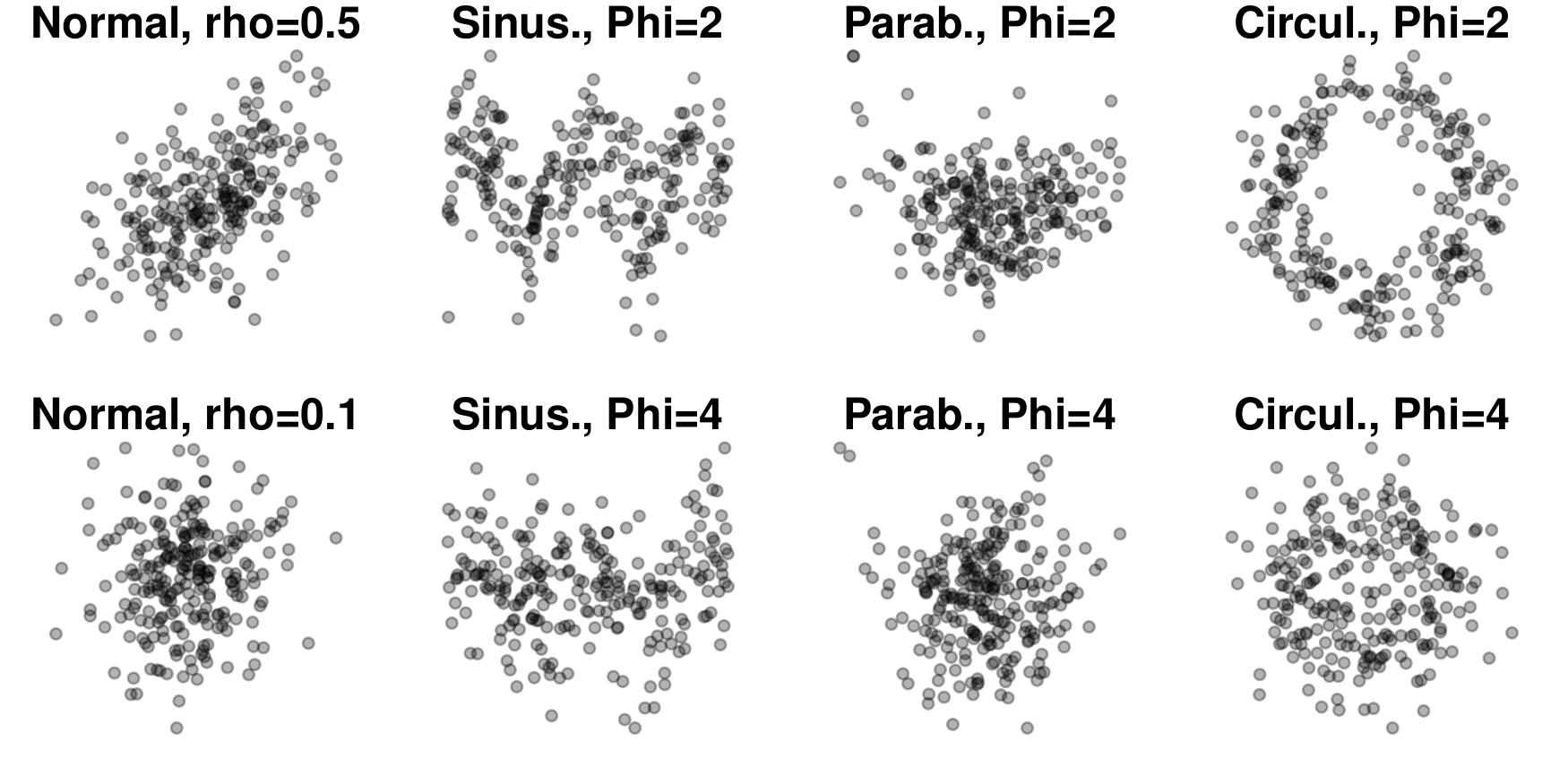}
\caption{Samples of size $250$ generated from the four scenarios for two levels of correlation $\rho$ in the normal model and two levels of noise $\phi$ in the sinusoidal, parabolic and circular models.}
\label{fig:models}
\end{figure}

For the sinusoidal, parabolic and circular models, the parameter $\phi$ controls the level of noise, whereas for the normal model the correlation $\rho$ controls the degree of dependence between the two samples. We generated fifty independent datasets from each model with a sample size $n=250$ with different correlations $\rho\in\{0,0.1,0.3,0.5,0.9\}$, for  model (a), and levels of noise $\phi\in\{1,2,3,4,5 \}$ for models (b)--(d). Figure~\ref{fig:models} shows one of the fifty simulated dataset as illustration. 

For all the simulated datasets we apply our different procedures for testing hypothesis \eqref{eq:m0m1}. We use the same priors specifications as described in Section~\ref{sec:WHO}.

\begin{figure}[H]
\includegraphics[width=\textwidth]{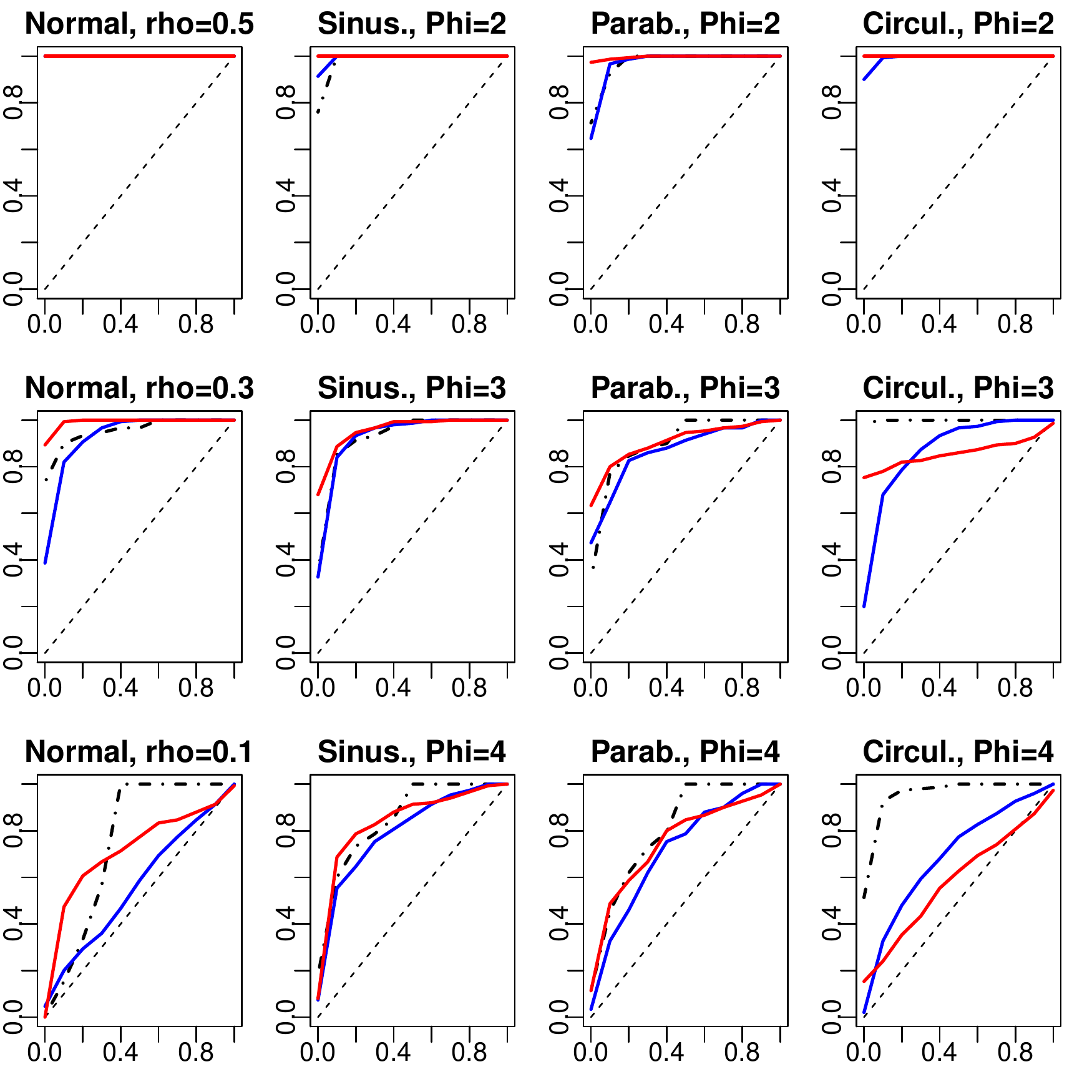}
\caption{ROC curves for competing methods as a function of correlation and noise level for models (a)--(d). CT-BF  (blue line), MixMod-ensemble (red line) and Mutual Information approximated using the $20$ nearest neighbours (black dotted line).} 
\label{fig:ROC_simp}
\end{figure}

To investigate the power of the two approaches, we create ROC curves that compare the rate of true positives (percentage of times the procedure detects dependence among the fifty datasets generated from a dependent model) and false positives (percentage of times the procedure detects dependence among fifty null datasets generated by randomly permuted the indexes of the two samples to destroy any dependences) for different threshold values. We also compare the performance of the proposed methods to the current state of the art conventional method, which is based on mutual information (using the $20$ nearest neighbours). The ROC curves are reported in Figure \ref{fig:ROC_simp}; see also \textit{Supplementary Material} that contains additional more extensive comparisons. 

We observe that the proposed methods have similar performances to the current leading conventional method
for data coming from a sinusoidal or a parabolic model. For data generated from the circular model however the mutual information method outperforms our approaches. 

%

\section{Conclusion}

We presented two Bayesian nonparametric procedures for highlighting pairwise dependencies between random variables that are scalable to large data sets. The methods make use of standard software in R for implementing DPM of Gaussians and are designed to exploit modern computer architectures. As such they are readily amenable to applied statisticians interested in exploratory analysis of large data sets. A power analysis shows that the procedures are comparable with that of current non-Bayesian methods based on mutual information, while having the advantage of being probabilistic in their measurement.

\label{sec:conclusion}


\bibliographystyle{abbrvnat} 
\bibliography{paper}

%
%
%
%
%
%
%

\newpage
{\huge Supplementary Material}
\appendix
\section{Comparison between variants of the two approaches}
In Section~\ref{sec:method} we described two Bayesian nonparametric approaches to highlight dependences between two random variables. For each approach, we mentioned different variants. Here we provide a comparison of these variants on the simulated dataset described in section~\ref{sec:simulated}. 

For the approaches based on contingency table, the main method consists in using all potential partitions coming from the Gibbs Sampling, performing the test at each iteration  and reporting the average probability of dependence (over all the iterations). An alternative to this approach, as mentioned in Section 3, is based on only one of the partitions selected using an optimisation criteria; we refer to this approach as CT-BF-1-clust. 

Regarding the mixture model approach, an alternative method to the posterior predictive approach of MiixMod-ensemble is a more conventional approach (called MixMod) described in algorithm~\ref{alg:MixModIt}. It consists in iteratively allocating the data to the independent or the dependent model and inferring each model based only on the data that has been allocated to it.

\begin{algorithm}               
\caption{Independence test MixMod}          
\label{alg:MixModIt}                           
\begin{algorithmic}                    
 \REQUIRE Data $\D=\{x_i,y_i\}_{i=1}^n$; Prior parameters $a_0$ and $b_0$; Prior parameters for the DPMs; Number of iterations $N_\text{it}$
    \ENSURE Estimate of mixture parameter $\pi$   
     \vspace{0.2cm}
     \STATE \underline{Initialisation:}
     \STATE $j\leftarrow 1$
     \STATE continue $\leftarrow$ true
    \STATE $\pi^{(1)}\leftarrow 0.5$
    \STATE $\xi\leftarrow$ vector with $n/2$ values equal to $0$ and $n/2$ values equal to $1$ randomly allocated
         \vspace{0.2cm}
    \WHILE{$ j\leq N_\text{it}$ and continue=true}
         \vspace{0.2cm}
 \STATE \underline{Infer DPMs given data allocation:}
    \STATE $n_\text{it}\leftarrow$ integer randomly sampled from $50$ to $100$
    \STATE $f_X^{(j)} \leftarrow$ posterior prediction of a DPM for distribution of $\{x_i, i \text{ s.t. } \xi_i=0\}$ based on the $n_\text{it}$-th  Gibbs sampler iteration
    \STATE $f_Y^{(j)} \leftarrow$ posterior prediction of a DPM for distribution of $\{y_i, i \text{ s.t. } \xi_i=0\}$ based on the $n_\text{it}$-th  Gibbs sampler iteration
    \STATE $f_{XY}^{(j)} \leftarrow$ posterior prediction of a DPM for joint distribution of $\{(x_i,y_i), i \text{ s.t. } \xi_i=1\}$ based on the $n_\text{it}$-th  Gibbs sampler iteration
         \vspace{0.2cm}
        \STATE \underline{Allocation of the data:}
        \FOR{$1\leq i\leq n$}
        \STATE   $p\leftarrow \left(\pi^{(j)}f_{XY}^{(j)}(x_i,y_i)\right)/\left(\pi^{(j)}f_{XY}^{(j)}(x_i,y_i)+(1-\pi^{(j)})f_X^{(j)}(x_i)f_Y^{(j)}(y_i)\right)$
        \STATE $\tilde\xi_i\sim \ber(p)$
        \ENDFOR
        \STATE $\tilde{l_0}\leftarrow \#\{\tilde\xi_i=0\}$
         \vspace{0.2cm}
       \IF{$\min(\tilde{l_0}, n-\tilde{l_0})\ge 5$}
       \STATE $\xi_i \leftarrow \tilde\xi_i$ for all $1\leq i\leq n$
       \STATE$l_0\leftarrow \#\{\xi_i=0\}$
       \ENDIF
        \vspace{0.2cm}
        \STATE $\pi^{(j+1)}\sim \be(a_0+l_0, b_0+n-l_0)$
             \vspace{0.2cm}

             \STATE $j\leftarrow j+1$
                  \ENDWHILE
         \vspace{0.2cm}
    
     \STATE $\hat{\pi}\leftarrow \frac{1}{N_\text{it}-N_\text{it}/10+1}\sum_{j=N_\text{it}/10}^{N_\text{it}}\pi^{(j)}$
    
\end{algorithmic}
\end{algorithm}

\begin{figure}
\includegraphics[width=\textwidth]{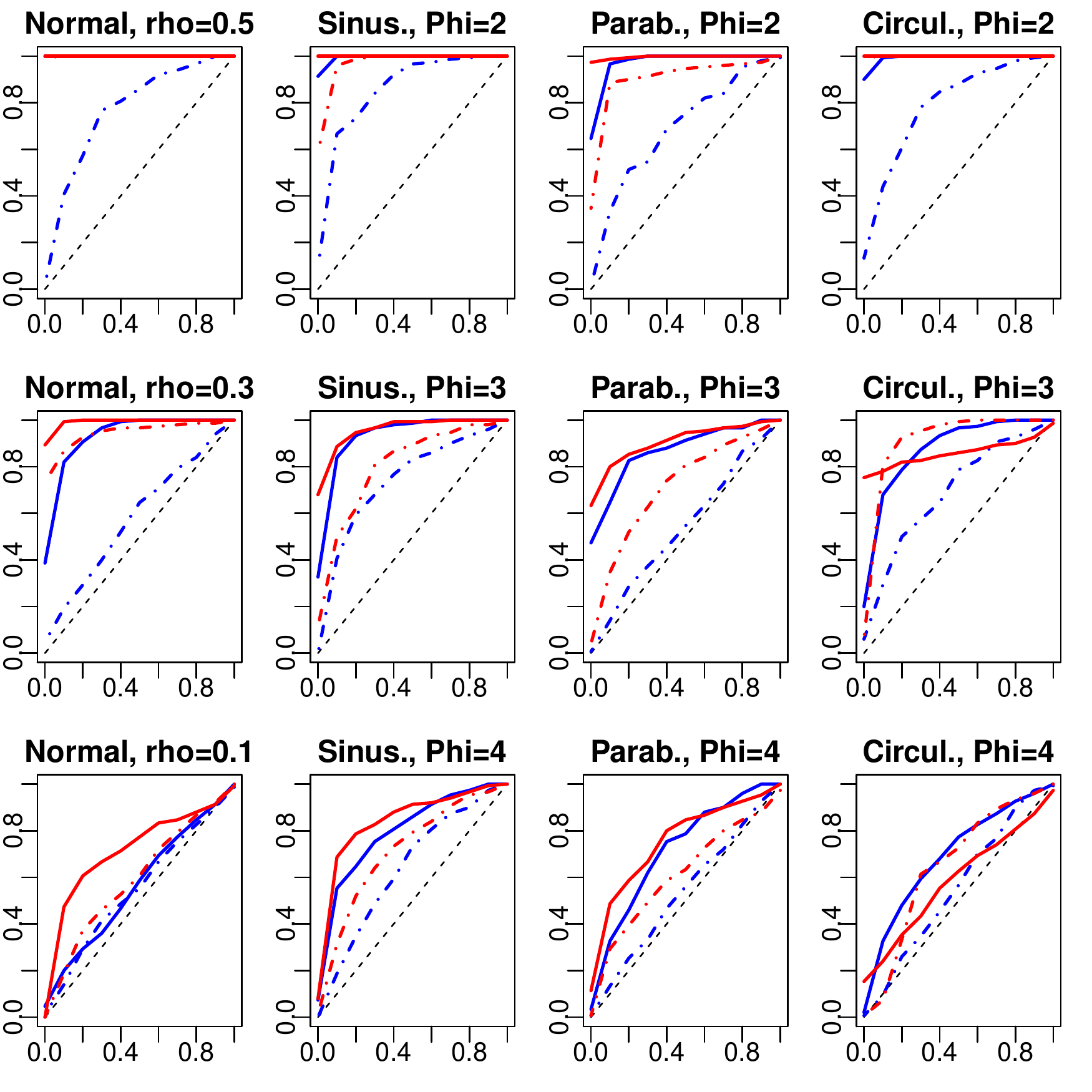}

\caption{ROC curves for competing methods as a function of correlation and noise level for models (a)--(d). CT-BF 1 clust (blue dashed line), CT-BF (blue solid line), MixMod (red dashed line), and MixMod-ensemble (red solid line).} 
\label{fig:ROC}
\end{figure}

We compare the four approaches (two alternatives for each approach) in terms of their frequentist power on the simulated examples of Section 4. The ROC curves are reported in Figure~\ref{fig:ROC}. As expected for all models, the statistical power increases for larger correlation values and decreases for larger levels of noise. We observe that the CT-BF approach using all iterations of the Gibbs sampling has significantly more power than the approach when using a unique cluster. Regarding the mixture models approach, the posterior predictive two steps approach is slightly more powerful and is computationally much cheaper than the iterative approach. 

\section{Sensitivity to prior choices}
All the proposed methods require the specification of several hyper-parameters controlling the prior distributions of the DPM and also of the test themselves. We have investigated the impact of these choices and observed that none of the approaches is sensitive to the choice of the hyper-parameters controlling $G_0$ or $G_1$ in the DPM models. In the following, we illustrate the impact of the other hyper-parameters on simulated data sampled from the sinusoidal model described in Section~\ref{sec:simulated}.

\begin{figure}
\includegraphics[width=\textwidth]{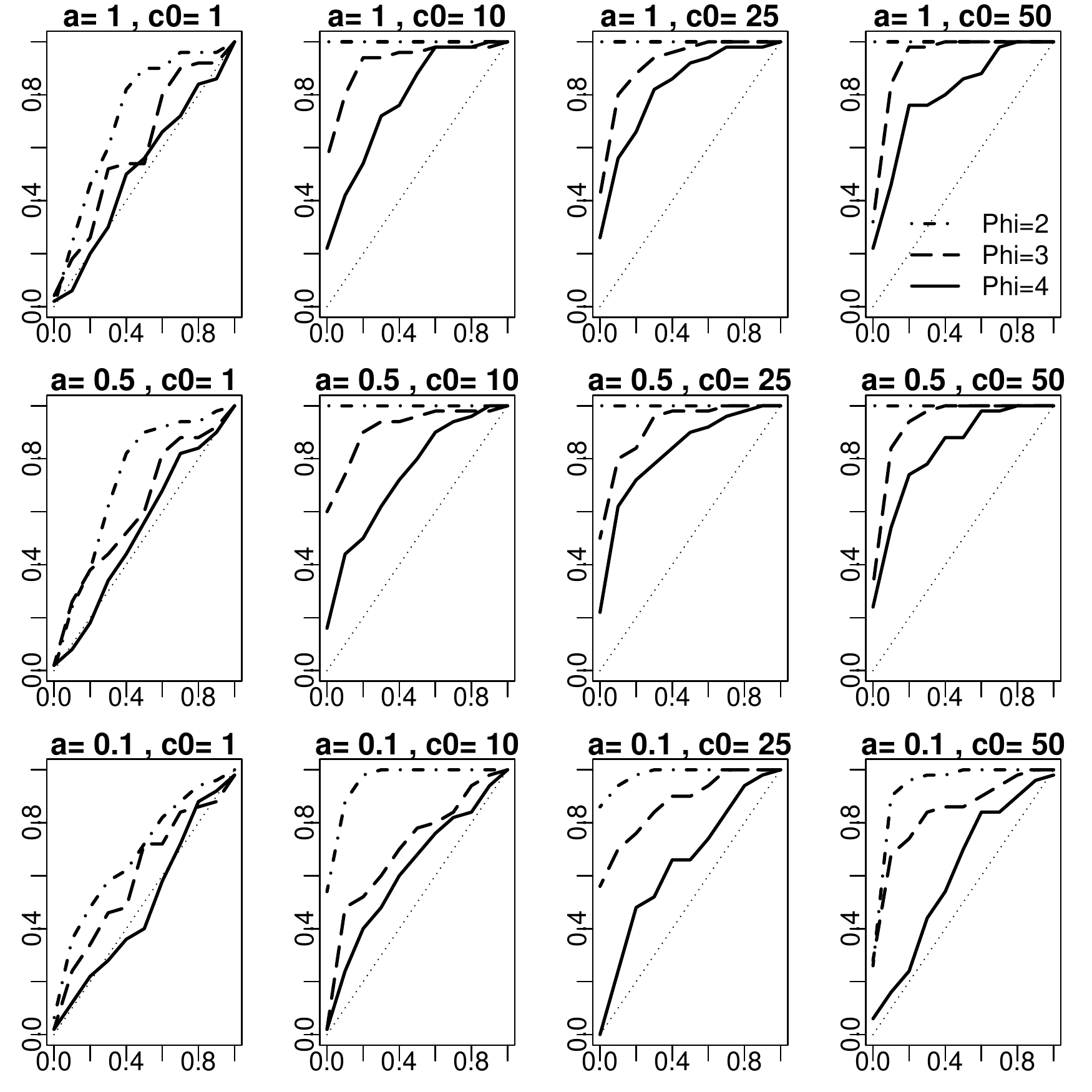}
\caption{Power analysis of the CT-BF method illustrated on the simulated sinusoidal example with different values of $\phi$ corresponding to different lines) varying the parameters $a$ and $c_0$. }
\label{fig:Sensitivity:CTROC}
\end{figure}

\begin{figure}
\includegraphics[width=\textwidth]{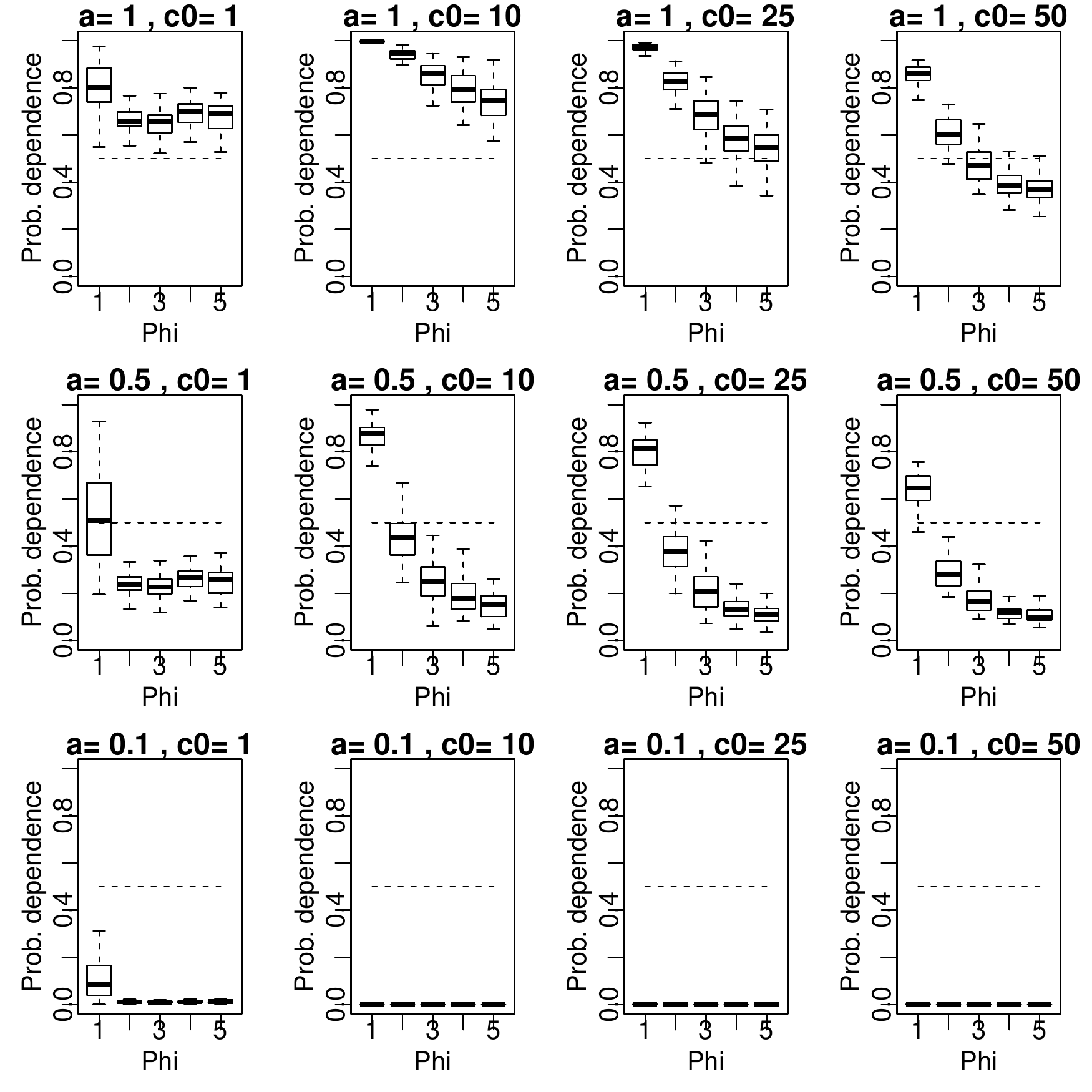}
\caption{Probability of dependence obtained following the CT-BF approach for different  simulated data generated from the sinusoidal example as a function of the noise level ($\phi$). Each panel corresponds to a different value of the parameters $a$ and $c_0$. The box plots illustrate the distribution of the probabilities obtained for 50 Monte-Carlo samples.}
\label{fig:Sensitivity:CT}
\end{figure}

The contingency table approach is highly sensible to the choice of the parameter $c_0$. This parameter influence the number of clusters in the DPM model and therefore the size of the contingency tables. The frequentist power of the method on the simulated sinusoidal example increases with the parameter $c_0$ (see Figure~\ref{fig:Sensitivity:CTROC}). However, the ROC curves are fairly insensitive to the choice of the parameters $a$ controlling the Dirichlet prior for the cell probability: $\alpha_{kl}=a$ for all $k$ and $l$. We suggest to use $a=0.5$ as  that value spreads the probability of dependence in the interval $(0,1)$ for the different levels of additive noise (see Figure~\ref{fig:Sensitivity:CT}).  

\begin{figure}
\includegraphics[width=\textwidth]{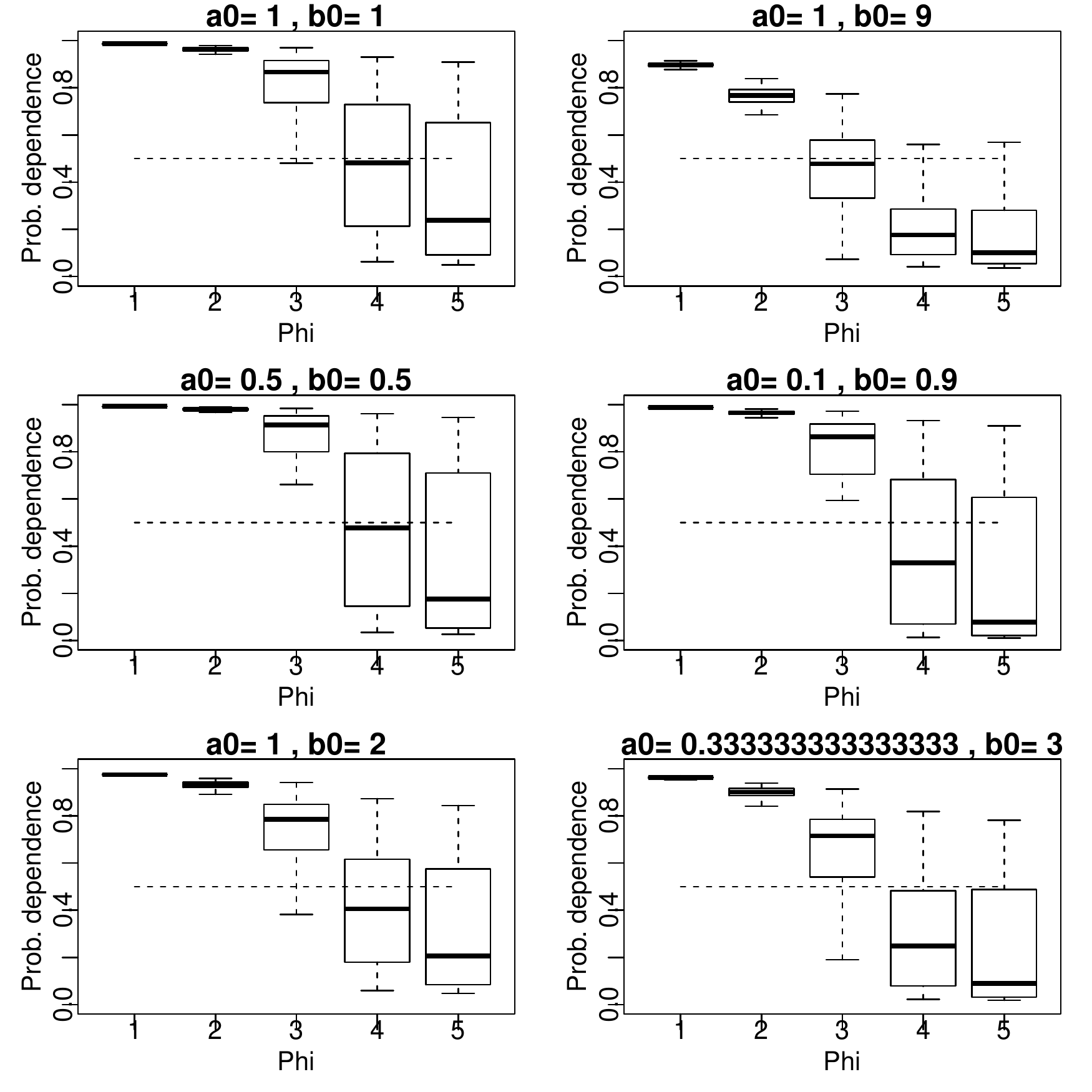}
\caption{Probability of dependence obtained following the MixMod-ensemble approach for different  simulated data generated from the sinusoidal example as a function of the noise level ($\phi$). Each panel corresponds to a different value of the parameters $a_0$ and $b_0$. The box plots illustrate the distribution of the probabilities obtained for 50 Monte-Carlo samples.}
\label{fig:Sensitivity:Mix}
\end{figure}

\begin{figure}
\includegraphics[width=\textwidth]{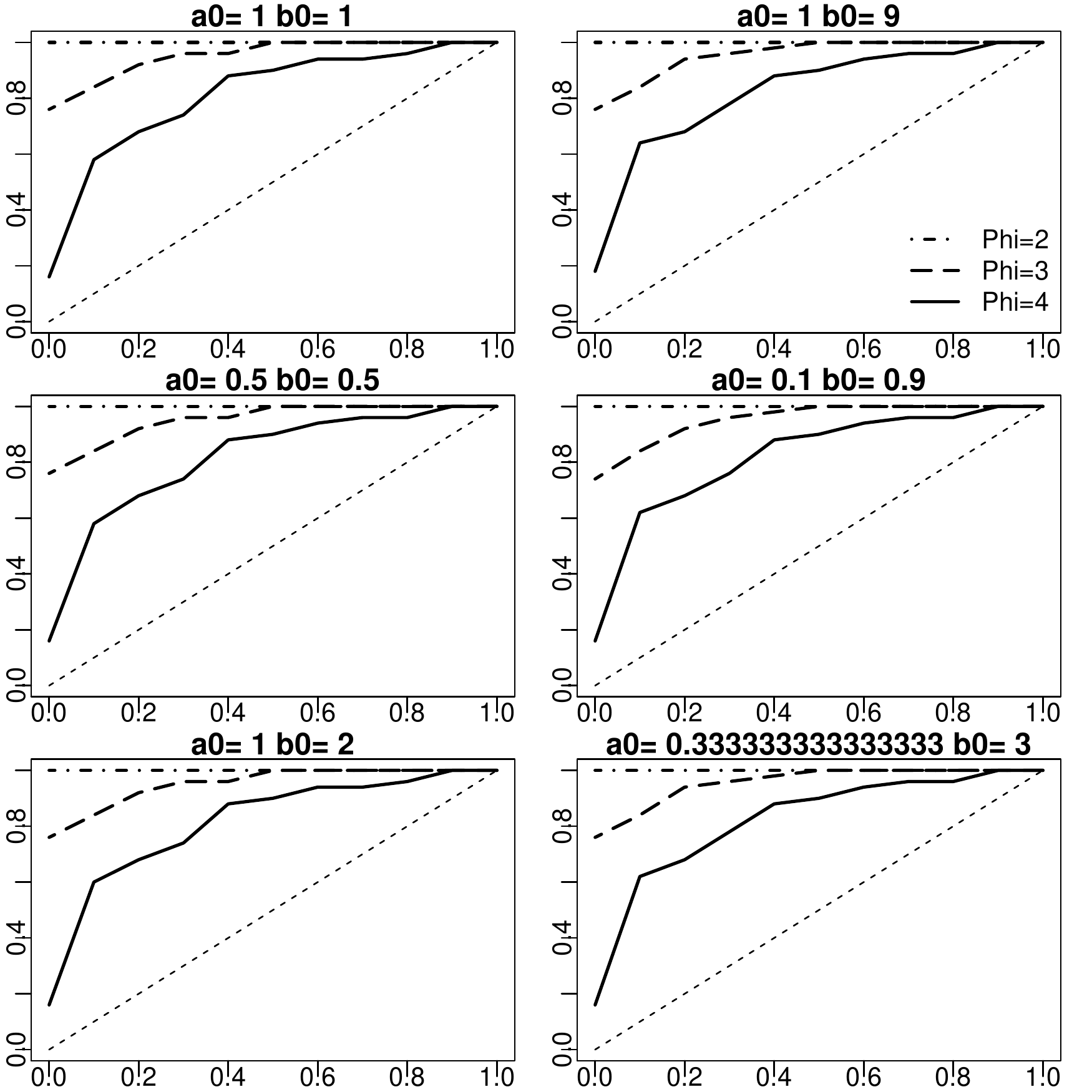}
\caption{Power analysis of the MixMod-ensemble method illustrated on the simulated sinusoidal example with different values of $\phi$ corresponding to different lines) varying the parameters $a_0$ and $b_0$. }
\label{fig:Sensitivity:MixROC}
\end{figure}

We observe in Figures~\ref{fig:Sensitivity:Mix} and~\ref{fig:Sensitivity:MixROC} that the performance of the MixMod-ensembl approach in term of power is not sensible to the choice of the parameters of the beta prior ($a_0$ $b_0$) and for the mixture proportion $\pi$.

%
%
\newpage
\section{R code}
In this section, we provide the R code that has been used to run our two independence tests on the WHO dataset. The data are stored in a matrix of size $194\times 562$, called \textit{data}. The first subsection present the file containing the main functions of interest and the second subsection shows how to call these routines in parallel.

\subsection{functions.R}
\begin{lstlisting}
ndata=dim(data)[2]

# Prior specifications for CTBF
gammavec = 10 
prior=list(alpha=gammavec,m2=rep(0,1),s2=diag(1,1),psiinv2=solve(diag(0.1,1)),nu1=3,nu2=3,tau1=1,tau2=100)

# Prior specifications for MixMod
prior_marg = list(a0=1,b0=10,m2=rep(0,1),s2=diag(100,1),
                   psiinv2=solve(diag(0.1,1)),
                   nu1=3,nu2=3,tau1=1,tau2=100)
prior_joint = list(a0=1,b0=10,m2=rep(0,2),s2=diag(100,2),
                    psiinv2=solve(diag(0.1,2)),
                    nu1=4,nu2=4,tau1=1,tau2=100)

# Information for MCMC scheme
state = NULL
mcmc = list(nburn=100,nsave=1000,nskip=5,ndisplay=10001)

# Function to fit DPMs on the two marginal and on the joint space
extractClusterInfoDPM_ind <-function(indvar){
  print(indvar)
  I=is.finite(data[,indvar])
  x=data[I,indvar]
  x=x-mean(x)
  x=x/sd(x)
  if (length(x)>10){
    id=seq(1,2*length(x),2,)
    tmp=tryCatch({
      x.fit = DPdensity(y=x,prior=prior,mcmc=mcmc,state=state,status=TRUE)
      tmp = matrix(NA,nsave,length(I))
      tmp[,I]=x.fit$save.state$randsave[,id]
      return(tmp)
    }, error=function(e){return(matrix(NA,nsave,length(I)))})
  }else{
    tmp=NA
  }
  return(tmp)
}


# Functions  to perform tests based on Contingency Tables
pFromAllEval=function(ind){
  i=indI[ind]
  j=indJ[ind]
  x.means.all=allEval[[i]]
  y.means.all=allEval[[j]]
  if(sum(is.finite(x.means.all))>1 & sum(is.finite(y.means.all))>1){
    I=!(is.na(x.means.all[1,])|is.na(y.means.all[1,]))
    if (length(which(I==T))>10){
      xhere=x.means.all[,I]
      yhere=y.means.all[,I]
      pH1=lapply(seq(nsave), function(k) pFromAllEval1iter(xhere,yhere,k))
      out =mean(as.numeric(pH1))
    }else{
      out=NA
    }
  }else{
    out=NA
  }
  return(out)
}


pFromAllEval1iter = function(x,y,k){
  x.means=round(x[k,],8)
  y.means=round(y[k,],8)
  m=table(y.means,x.means)
  kx=ncol(m)
  ky=nrow(m)
  # computing Dirichlet test
  a=0.5	
  mx=apply(m,2,sum)
  my=apply(m,1,sum)
  bm=sum(lgamma(a+m))-lgamma(sum(a+m))
  bmx=sum(lgamma(ky*a+mx))-lgamma(sum(ky*a+mx))
  bmy=sum(lgamma(kx*a+my))-lgamma(sum(kx*a+my))
  b=sum(lgamma(a+m-m))-lgamma(sum(a+m-m))
  bx=sum(lgamma(ky*a+mx-mx))-lgamma(sum(ky*a+mx-mx))
  by=sum(lgamma(kx*a+my-my))-lgamma(sum(kx*a+my-my))
  bf=exp(bmx-bx+bmy-by+b-bm)
  p1=1/(1+bf)
  return(p1)
}


# Function to perform MixMod-ensemble test
MixMod2steps = function(ind){
  x=data[,indI[ind]]
  y=data[,indJ[ind]]
  I=!(is.na(x)|is.na(y))
  x=x[I]
  y=y[I]
  x=x-mean(x)
  y=y-mean(y)
  x=x/sd(x)
  y=y/sd(y)
  # do the two tests
  if (length(x)>10){
    x.fit = DPdensity(y=x,prior=prior_marg,mcmc=mcmc2,
                       state=state,status=TRUE, grid=x)
    y.fit = DPdensity(y=y,prior=prior_marg,mcmc=mcmc,
                       state=state,status=TRUE, grid=y)
    xy.fit = DPdensity(y=cbind(x,y),prior=prior_joint,mcmc=mcmc2,
                        state=state,status=TRUE, grid=cbind(x,y))
    a0=0.5
    b0=0.5          
    xy_d=diag(xy.fit$dens)
    interv=0.0001
    alphaGrid = seq(interv,1-interv,interv)
    logpa =rep(NA,length(alphaGrid))
    for (it in 1:length(alphaGrid)){
      alpha=alphaGrid[it]
      logpa[it]=sum(log(alpha*xy_d+(1-alpha)*x.fit$dens*y.fit$dens))+log(dbeta(alpha,a0,b0))
    }
    pa=exp(logpa+min(-500-logpa))/sum(exp(logpa+min(-500-logpa))*interv)	
    res=sum(pa*alphaGrid)/length(alphaGrid)	
  }else{
    res=NA
  }
  return(res)	
}
\end{lstlisting}

\subsection{main.R}
\begin{lstlisting}
library(parallel)
library(DPpackage)
source("functions.R")

ncore=48

# Construct list of indexes
ntmp=matrix(NA,ncol=ndata,nrow=ndata)
indI=matrix(seq(ndata),ncol=ndata,nrow=ndata)[upper.tri(ntmp)]
indJ=matrix(seq(ndata),ncol=ndata,nrow=ndata, byrow=T)[upper.tri(ntmp)]

######### For the CT-BF approach

# for each variable, run DPM
cl = makeCluster(ncore)
clusterEvalQ(cl, source("functions.R"))
allEval=parLapply(cl, seq(ndata), function(ind) extractClusterInfoDPM_ind(ind))
stopCluster(cl)

#Run all the tests in parallel
cl = makeCluster(ncore)
clusterEvalQ(cl, source("functions_parallelCTBF.R"))
clusterExport(cl, c("allEval", "indI","indJ"))
pH1=parLapply(cl,seq(length(indI)), function(k) pFromAllEval(k))
stopCluster(cl)

# reshape the results and write them in a text file
pCT=matrix(NA,ndata,ndata)
pCT[upper.tri(pCT)]=as.numeric(pH1)
write.table(pCT,"WHOres_pCT_parallel.txt")

######### For the MixMod-ensemble approach

cl = makeCluster(ncore) 
clusterEvalQ(cl, source("functions.R"))
clusterExport(cl, c("indI","indJ"))
t1=Sys.time()
pH1=parLapply(cl,seq(length(indI)),function(i) tryCatch({tmp=MixMod2steps(i)}, error=function(e){return(NA)}))
print(Sys.time()-t1)
stopCluster(cl)

# reshape the results 
pMM=matrix(NA,ntot,ntot)
pMM[upper.tri(pMM)]=as.numeric(pH1)
write.table(pMM,"WHOres_pMixMod_parallel.txt")

\end{lstlisting}

\end{document}